\documentclass[conference,a4paper]{IEEEtran}
\usepackage[latin1]{inputenc}

\usepackage{makeidx}
\usepackage{verbatim}
\usepackage{amsmath}
\usepackage{amsfonts}
\usepackage{amssymb}

\newtheorem{defi}{Definition}

\newtheorem{prop}{Proposition}
\newtheorem{thm}{Theorem}

\newtheorem{rmq}{Remark}

\newtheorem{xpl}{Example}

\newcommand{\poubelle}[1]{}

\newcommand{\Tordu}[2]{#1^{\theta^{#2}}}
\newcommand{\Gal}{\operatorname{Gal}}
\newcommand{\rank}{\operatorname{rank}}

\newcommand{\F}{{\mathbb F}}
\newcommand{\Diag}{\operatorname{Diag}}

\newcommand{\refeq}[1]{Eq.~\ref{#1}}
\newcommand{\floor}[1]{\lfloor#1\rfloor}
\newcommand{\C}{\mathbb C}
\newcommand{\Q}{\mathbb Q}
\newcommand{\Z}{\mathbb Z}
\usepackage{amsmath}
\newcommand{\defeq}{\stackrel{\text{def}}{=}}

\newcommand{\MatriceCoins}[4]{
\left( \begin{array}{ccc}
    #1 & \cdots & #2 \\ 
    \vdots & \ddots & \vdots \\ 
    #3 & \cdots & #4 \\ 
  \end{array} \right) 
}

\newcommand{\VecteurLigneCoins}[2]{
\left( #1, \cdots,  #2  \right) 
}

\newcommand{\MatriceGeneraleEch}[3]{
\left( \begin{array}{ccc}
    #1_{1,1} & \cdots & #1_{#3,1} \\ 
    \vdots &  \ddots & \vdots \\ 
    #1_ {1,#2} & \cdots & #1_{#3,#2} \\ 
  \end{array} \right) 
}

\begin{document}

\title{Rank metric and  Gabidulin codes in characteristic zero}
\author{
\IEEEauthorblockN{Daniel Augot}
  \IEEEauthorblockA{
    INRIA Saclay-\^Ile-de-France\\
    \'Ecole polytechnique\\
    Palaiseau, France\\
    Email: Daniel.Augot@inria.fr}
  \and
  \IEEEauthorblockN{Pierre Loidreau}
  \IEEEauthorblockA{DGA and IRMAR \\ 
    Universit\'e de Rennes 1\\
    Rennes, France\\
    Email: Pierre.Loidreau@univ-rennes1.fr}
\and
  \IEEEauthorblockN{Gwezheneg ROBERT}
  \IEEEauthorblockA{IRMAR\\
    Universit\'e de Rennes 1\\
    Rennes, France\\
    Email: Gwezheneg.Robert@univ-rennes1.fr} 
}

\maketitle

\begin{abstract}
  We transpose the theory of rank metric and Gabidulin codes to the
  case of fields of characteristic zero.  The Frobenius automorphism
  is then replaced by any element of the Galois group.  We derive some
  conditions on the automorphism to be able to easily transpose the
  results obtained by Gabidulin as well and a classical
  polynomial-time decoding algorithm. We also provide various
  definitions for the rank-metric.
\end{abstract}


\section{Motivation}

Matricial codes with coefficients in a finite subset of the complex
field are particularly well-suited for the design of space-time
codes. When the metric of the code space is the rank metric, its
minimum distance is called the diversity.  This parameter is one of
the crucial parameters in evaluating the performance of Minimum
Distance Decoding \cite{Hammons/ElGamal:2000}.

A problem in the field of space-time coding is to construct codes with
optimal rate/diversity trade-off.  Lu and Kumar~\cite{Lu/Kumar:2005}
used an original approach by transforming optimal codes in rank metric
over finite fields, such as Gabidulin codes, into optimal codes for
space-time coding over different types of constellations.

%MODIF
However a mapping $\mathbb{F}_q^k \rightarrow \mathbb{C}$ is
used, which is difficult to reverse, yet its inverse is needed to
recover information bits when decoding.
%fin de MODIF
Another construction based on Gabidulin codes over finite fields has
been given in~\cite{Lusina2003}, using particular properties of
Gaussian integers.

We propose in this paper to construct optimal codes similar to
Gabidulin codes, with coefficients in $\mathbb C$, completely
bypassing intermediate constructions using finite fields, using number
fields and Galois automorphims. We also provide a decoding algorithm
using with a polynomial number of field operations (this is not the
bit complexity). 

Further work is needed to study the proper use of this construction in
the area of space-time coding.

%If it were possible, then some
%immediate consequences would happen:
%
%\begin{itemize}
%\item Any subset of these codes would have a diversity at least equal
%  to that of the code. Therefore, one could try to extract subcodes
%  with good properties. They might even be linear codes over some
%  field, whereas the constructions by Lu and Kumar constructs
%  essentially non-linear codes.
%\item The code itself comes directly with a decoding algorithm up to
%  half the diversity. This algorithm would have a polynomial algebraic
%  complexity. This could be of interest, since usually space-time
%  codes are decoded by enumerating all the codewords and taking the
%  nearest one (for a well-defined metric) to the received vector.
%\end{itemize}
%
%Phrase qui a été enlevée pour la version ISIT :
%However, we do not consider finite fields any more, but number fields,
%where the Galois groups can be much more complicated.

\section{Contribution}

% In this paper, we mimic the construction of Gabidulin codes and the
% study of their properties in a field of characteristic zero.

In the original paper of Gabidulin, the constructed codes are
evaluation codes of linearized
polynomials~\cite{Lidl-Niederreiter:FF1996} with coefficients in a
finite field. The associated metric is called \emph{rank metric} and
is of interest for correcting errors which occur along rows or columns
of matrices.  Transposing the results in characteristic zero fields is
more tricky. Namely, in finite fields the Galois groups are well known
and the field extensions are all cyclic. However in characteristic
zero, it is abolutely not the case and one needs to be very careful
and find some criteria so that we can transpose Gabidulin construction
in that case.

We call polynomials equivalent to linearized polynomials \emph{$\theta$-polynomials}, where $\theta$ is an automorphism of a field extension $K \hookrightarrow L$ of degree $m$. The automorphism $\theta$ is of order $n$, which divides $m$. In the first section we establish conditions 
%MODIF condition au lieu de propriete, ce mot étant utilisé dans la suite de la phrase.
such that the $\theta$-polynomials present robust properties, namely that the root-space of a $\theta$-polynomial has dimension less than its degree. In a second section, we show that all the different possible metrics that we could think of concerning rank metric are in fact the same provided that the base field is exactly the fixed field of $\theta$. Under this condition, we can define the {\em rank metric} in a unique way.

In the final section we construct Gabidulin codes, showing that they are optimal for the rank metric and that they can be decoded by using some of the existing decoding algorithms. And finally we give some examples. We refer the reader to \cite{Lang} for basics on Galois theory.

%%% Local Variables: 
%%% mode: latex
%%% TeX-master: "Soumission_ISIT"
%%% End: 

\section{$\theta$-polynomials}
\label{Section:ThetaPoly}
In all the paper, we consider an algebraic field extension $K
\hookrightarrow L$ with finite degree $m$, and an automorphism
$\theta$ in the Galois group $\Gal( K \hookrightarrow L )$, of order
$n\leq \Gal( K \hookrightarrow L )\leq m$.  Given $v \in L$, we use
the notation $v^{\theta^i}$ for $\theta^i(v)$.
%MODIF pour expliquer la notation:
In the finite field case, when $\theta$ is the Frobenius automorphism
$x \mapsto x^q$, $\Tordu{v}{i}=v^{q^i}$, and the similarity is nicely
reflected in the notation.

We note $\mathcal{B}\defeq(b_1,\ldots,b_m)$ a $K$-basis of
$L$. For finite fields, we use the notation $\F_q \hookrightarrow
\F_{q^m}$. Similarly to \emph{linearized polynomials}, we define
\emph{$\theta$-polynomials}, which is a special case of \emph{skew polynomials}, namely, when there is no derivation.
%MODIF
\begin{defi} A $\theta$-polynomial is a finite summation of the form
  $\sum_i p_i \Tordu{X}{i}$, with $p_i \in L$.  The greatest integer
  $i < \infty$ such that $p_i \ne 0$ is called its $\theta$-degree,
  and is denoted by $\deg_{\theta}(P)$. 
\end{defi}
We denote the set of $\theta$-polynomials by $L[X ; \theta]$. We have
the following operations on the set $L[X;\theta]$:
\begin{enumerate} 
\item Componentwise scalar multiplication and addition;
\item Multiplication: for  $P(X) = \sum_{i} p_i
  \Tordu{X}{i}$ and $Q(X) = \sum_{i} q_i \Tordu{X}{i}$, 
\[
 P (X) \cdot Q (X) = \sum_{i,j} p_i\, \Tordu{q_j}{i} \Tordu{X}{i+j};
\]
\item Evaluation: Given $v\in L$, and $P(X) = \sum_{i} p_i
  \Tordu{X}{i}$:
\[
P(v)=\sum p_i \Tordu{v}{i}.
\]
\end{enumerate}
%MODIF pour préciser d'où vient la formule du produit :
The multiplication formula is motivated by the composition law: $P(X)
\cdot Q(X) = P(Q(X))$.  The following is well known.
\begin{prop}[\cite{Ore:1932}]
  The set of $\theta$-polynomials $(L[X ; \theta], + , \cdot)$ is a
  non-commutative integral domain, with unity $\Tordu{X}{0}$. 
  It is also a left and right Euclidean ring.
\end{prop}
Such a ring is an Ore ring with trivial derivative. The proof is the same regardless of the characteristic of the fields.
%MODIF le rapporteur E utilise le mot derivation, alors que Ore utilisait derivative. 
Considering the case where $K = \mathbb{F}_q $ and $L =
\mathbb{F}_{q^m}$ are finite fields, and where $\theta$ is the
Frobenius automorphism $x \mapsto x^q$,  we get the set of
\emph{linearized polynomials}, also called \emph{$q$-polynomials}. In
that particular case, one has the following important proposition.
\begin{prop}[\cite{Ore:1933}]\label{label:ore33}
  The roots of a $q$-polynomial with $q$-degree $t$ form a
  $\mathbb{F}_q$-vector space with dimension at most $t$.
\end{prop}
We define the root-space of a $\theta$-polynomial $P(X)$ to be the set
of $v\in L$ such that $P(v)=0$.  Then Prop.~\ref{label:ore33}
does not generalize to more general $\theta$-polynomials, when
$\theta$ is not well behaved, as shown below.

\begin{xpl}\label{Ex:Racines}
  Here is an example of a $\theta$-polynomial whose root-space
  dimension is twice its $\theta$-degree.  Let us consider the field
  extension
\[
K = \mathbb{Q} \hookrightarrow L=\mathbb{Q}[Y]/(Y^8+1).
\]
Let $\alpha$ be a root of $Y^8+1$, such that $(1, \ldots , \alpha^7)$ is a $K$-basis of $L$. 
Consider the automorphism $\theta$ defined by $ \alpha \mapsto
\alpha^3$. 
The polynomial $\Tordu{X}{1}-\Tordu{X}{0}$ has a root-space of dimension 2, with two
$K$-generators: $1$ and $\alpha^2+\alpha^6$.
One can actually check that the characteristic polynomial of $\theta$ as a $K$-linear map is
$Y^8-2Y^4+1 = (Y^4-1)^2$, i.e. non square-free, which is the cause of the problem.
\end{xpl}

%\begin{xpl}
%Here is an example of a $\theta$-polynomial whose root-space dimension can reach twice its $\theta$-degree.
%Let us consider the Kummer extension
%\[
%K = \mathbb{Q}[X]/(X^4+1) \hookrightarrow L=K[Y]/(Y^8-3).
%\]
%Let $h$ be a root of $X^4+1$, such that $(1,h,h^2,h^3)$ is a $\mathbb{Q}$-basis of $K$, and let $\alpha$ be a root of $Y^8-3$, such that $(1,\ldots,\alpha^7)$ is a $K$-basis %of $L$.
%Consider the automorphism $\theta$ defined by $ \alpha \mapsto h^2 \alpha$. 
%One can check that its characteristic polynomial of $\theta$ (see as a linear map) is $Y^8-2Y^4+1=(Y^4-1)^2$.
%Thus, the polynomial $X^{\theta^1}-X^{\theta^0}$ has a root-space of dimension 2, with two $K$-generators: $1$ and $\alpha^4$.
%\end{xpl}

Thus we have a simple criteria on $\theta$ to establish a property
equivalent to Prop.~\ref{label:ore33} in the general case.
\begin{thm}\label{theo:root}
  If the characteristic polynomial of $\theta$, considered as a
  $K$-linear application, is square-free, then the dimension of the
  root-space of a $\theta$-polynomial is less than or equal to its
  $\theta$-degree.
\end{thm}
\begin{IEEEproof}
  Let $P(X) = \sum p_i \Tordu{X}{i}$. Let us denote
  $\overline{P}(X)\defeq\sum p_i X^i$.  Let $M$ be the matrix of $\theta$
  in the basis $\mathcal{B}$. Let $y$ be an element of $L$ and
  $Y_{\mathcal{B}}$ the $m$-dimensional vector in $K^m$ corresponding
  to its representation in the basis $\mathcal{B}$.  We have
\[
P(y) \defeq\sum_i {p_i \theta^i(y)} = \overline{P}(M)
\cdot Y_{\mathcal{B}}.
\] 
Therefore the root-space of $P$ is equal to the right kernel of the
matrix $\overline{P}(M)$.  Since by hypothesis the characteristic
polynomial of $\theta$ is square-free, all its roots are distinct.
Let $\alpha_1,\cdots,\alpha_m$ be its roots. Since $\theta$ is
invertible $0$ is not a root of the polynomial.  Therefore, there
exists a $ m \times m$- non-singular matrix $Q$ with coefficients in
$K$, such that $M=Q^{-1} \cdot \underbrace{\Diag \VecteurLigneCoins{\alpha_1}{\alpha_m}}_D \cdot Q$.  Hence
\begin{IEEEeqnarray*}{rll}
\overline{P}(M) &= Q^{-1} \cdot \sum_i {p_i D^i} \cdot Q \\
{} &=  Q^{-1} \cdot \Diag \VecteurLigneCoins{\overline{P}(\alpha_1)}{\overline{P}(\alpha_m)} \cdot Q.
\end{IEEEeqnarray*}
Therefore the dimension of the root-space of $P$ is equal to the
number of $\alpha_i$'s which are roots of $\overline{P}$. Since by
hypothesis the $\alpha_i$'s are distinct and since the degree of
$\overline{P}$ is the same as the degree of $P$, the dimension of the
root-space of $P$ is at most its degree.
\end{IEEEproof}
Note that the condition that the characteristic polynomial is
square-free implies that $K=L^\theta$. We also need the following
theorem, which show that we can find annihilator polynomials of
$K$-subspaces of $L$.
\begin{thm} 
\label{Thm:Lagrange}
Let $\theta$ have a square-free characteristic polynomial. Let
$\mathcal{V}$ be an $s$-dimensional $K$-subspace of $L$. Then there
exists a unique monic $\theta$-polynomial $P_{\mathcal{V}}$ with
$\theta$-degree $s$ such that
 \begin{equation}\label{poly:"candidate"}
  \forall v \in \mathcal{V},\quad P_{\mathcal{V}}(v) = 0.
 \end{equation}
\end{thm}
\begin{IEEEproof}
  The result is proven by induction.  Suppose first that $\mathcal V$
  has dimension 1, with $\mathcal{V} = \langle v_1\rangle$, where
  $v_1$ is non-zero element of $L$. Then $P_{\mathcal{V}} =
  \Tordu{X}{1} - \frac{\theta(v_1)}{v_1} \Tordu{X}{0}$
  satisfy~\refeq{poly:"candidate"}.  Suppose now that $\mathcal V$ has
  dimension $i+1$, with $\mathcal{V} = \langle
  v_1,\ldots,v_{i+1}\rangle$. The vectorspace $\mathcal{V}' = \langle
  v_1,\ldots,v_i\rangle$ has dimension $i$ and
$$
P_{\mathcal{V}}(X) = \left( X^{\theta^1} -
  \frac{\theta(P_{\mathcal{V'}}(v_{i+1}))}{P_{\mathcal{V'}}(v_{i+1})}\Tordu{X}{0}
\right) \times P_{\mathcal{V'}}(X)
$$ 
can be checked to satisfy~\refeq{poly:"candidate"}. It is monic and
has $\theta$-degree $i+1$.  Nevertheless we need to ascertain that
$P_{\mathcal{V}'}(v_{i+1}) \ne 0$: Since by hypothesis the root-space
of $P_{\mathcal{V}'}$ has dimension less than its degree and since
$v_{i+1}$ is not in this root-space, we get the desired result. To prove unicity, consider two monic $\theta$-polynomials $P_{\mathcal
  V}$ and $Q_{\mathcal V}$ and of degree $s$ vanishing on $\mathcal
V$. Then $P_{\mathcal V}-Q_{\mathcal V}$ has degree less than $s$ and
admits $\mathcal V$ among its roots. This contradicts
Th.~\ref{theo:root}.
\end{IEEEproof}

%%% Local Variables: 
%%% mode: latex
%%% TeX-master: "Soumission_ISIT"
%%% End: 

\section{Rank metric} In this section we present four definitions for
the rank weight. We show that in fact they define only two different
weights.  We also give a condition under which these two weights are
equal.
\begin{defi} Let $X =\VecteurLigneCoins{x_1}{x_N} \in L^N$. We define
  \[X^{\theta}\defeq
\MatriceCoins{\Tordu{x_1}{0}}{\Tordu{x_N}{0}}{\Tordu{x_1}{n-1}}{\Tordu{x_N}{n-1}},
\] and
\[ X_{\mathcal{B}} \defeq \MatriceGeneraleEch{x}{m}{N},
\] where $x_i= \sum_{j=1}^{m} x_{i,j} b_j$. We also define the left
ideal
\[ I_X \defeq \left\{ P \in L[X ; \theta] : P(x_i)=0, \;
i=1,\ldots,N\right\}.
\] The ideal $I_X$ being a left ideal in a right Euclidian ring, it
admits a right generator, denoted by $\min(I_X)$.

For any $X \in L^N$, we define the following quantities:
\begin{itemize}
\item $w_0(X) \defeq \deg_{\theta}(\min(I_X))$ ;
\item $w_1(X) \defeq {\rank}_L \left( X^{\theta} \right) =
\rank\left( X^{\theta} \right)$;
\end{itemize} which are related to $L$-linear independance, while the
following definitions are related to $K$-linear independance:
\begin{itemize}
\item $w_2(X) \defeq \rank_K \left( X^{\theta} \right) $;
\item $w_3(X) \defeq \rank_K \left( X_{\mathcal{B}} \right)
= \rank\left( X_{\mathcal{B}} \right)$;
\end{itemize} where $\rank_K$ stands for the maximum numberof $K$-linearly
independent columns.
\end{defi}
\begin{prop}
For all $X \in L^N$, $w_0(X)=w_1(X)$.
\end{prop}
\begin{IEEEproof}
  Let us denote $w_0(X)=w_0$, $w_1(X)=w_1$.  Since $\min(I_X)$ has
  degree $w_0$, then for any non zero $\VecteurLigneCoins{c_1}{c_{w_0-1}} \in L^N
  $, we have
\[
\VecteurLigneCoins{c_0}{c_{w_0-1}} \cdot
\MatriceCoins{\Tordu{x_1}{0}}{\Tordu{x_N}{0}}{\Tordu{x_1}{w_0-1}}{\Tordu{x_N}{w_0-1}} \ne 0.
\]
Thus the row rank over $L$ of $X^{\theta}$ is larger than or equal to
$ w_1$. Therefore $w_0 \le w_1$.

Writing   $\min(I_X) = \sum _{k=0}^{w_0} a_i \Tordu{x}{i}$, we have
\[
\VecteurLigneCoins{a_0}{a_{w_0}} \cdot 
\MatriceCoins{\Tordu{x_1}{0}}{\Tordu{x_N}{0}}{\Tordu{x_1}{w_0}}{\Tordu{x_N}{w_0}} = 0.
\]
Therefore the $(w_0+1)$-th row of $X^{\theta}$ is a $L$-linear
combination of the $w$ first rows of $X^{\theta}$. Applying $\theta^i$,
we have for all $i$:
\[
\VecteurLigneCoins{\Tordu{a_0}{i}}{\Tordu{a_w}{i}} \cdot 
\MatriceCoins{\Tordu{x_1}{i}}{\Tordu{x_N}{i}}{\Tordu{x_1}{w+i}}{\Tordu{x_N}{w+i}} = 0.
\]
This implies that the $(r+i)$-th row is a $L$-linear combination of
the $w_0$ preceeding rows,  thus of the $w_0$ first rows, by
induction.  Thus the $L$-rank of $X^{\theta}$ is less than $w_0$, and
$w_1 \le w_0 $.
\end{IEEEproof}
\begin{prop}
For all $X\in L^N$,  $w_2(X)=w_3(X)$.
\end{prop}
\begin{IEEEproof}
  Let $w_3=w_3(X)=\rank_K \left( X_{\mathcal{B}} \right) $, and
  $w_2=w_2(X)=\rank_K \left( X^{\theta} \right) $. Without loss of
  generality, suppose that the first $w_3$ columns of $X_\mathcal{B}$
  are $K$-linearly independent.  Accordingly, consider the $w_3$ first
  columns of $X^{\theta}$, and suppose that we have a dependence
  relation among them, i.e.
\[
\sum_{i=1}^{w_3}{\lambda_i \Tordu{x_i}{j}} = 0, \quad j=0,\ldots,n-1,
\]
with $\VecteurLigneCoins{\lambda_1}{\lambda_{w_3}} \in K^{w_3}$. Considering only
$j=0$, and rewriting $x_i=\sum_{j=1}^m{x_{i,j}}b_j$ over the basis
$\mathcal{B}$, we get
\[
0=\sum_{i=1}^{w_3}{\lambda_i \sum_{j=1}^m{x_{i,j}}b_j}=
\sum_{j=1}^{m}{\left(\sum_{i=1}^{w_3}{\lambda_i x_{i,j}}\right)b_j}.
\] Since the $b_i$'s are a $K$-basis, we have, for $j=1,\ldots,m$,
\[
0=\sum_{i=1}^{w_3}{\lambda_i x_{i,j}} =
X_\mathcal{B}  \cdot {\VecteurLigneCoins{\lambda_1}{\lambda_{w_3}}}^T.
\] 
By hypothesis the first $w_3$ columns of $ X_\mathcal{B}$ are linearly
independent, this implies $\lambda_i=0$, $i=1,\ldots,w_3$.  So the
first $w_3$ columns of $X^{\theta}$ are $K$-linearly independent.
Therefore $w_2\ge w_3$.

To prove that $w_2\leq w_3$, let $(x_{i,j})_{j=1}^m$ be the $i$-th
column of $X_\mathcal{B}$.  Since the first $w_3$ columns of
$X_\mathcal{B}$ generate the column space, we have $ x_i =
\sum_{k=1}^{w_3}{\lambda_{iu} x_{u}}$, $i=1,\ldots,N$. By $K$-linearity of
$\theta^j$, we have
\[
\Tordu{x_i}{j} = \sum_{u=1}^{w_3}{\lambda_{iu} \Tordu{x_u}{j}}, \quad
j=0,\ldots,n-1,
\]
therefore the $i$th column $(\Tordu{x_i}{j})_{j=0}^{n-1}$ of $X^{\theta}$ is generated by the first $w_3$ columns of $X^{\theta}$, and $w_2\le w_3$.
\end{IEEEproof}
\begin{prop}
  For all $X\in L^N$, $w_1(X) \leq w_2(X)$, with equality when $K  $ is
  the fixed subfield of $L$, i.e.\ $K=L^\theta$.
\end{prop}
\begin{IEEEproof}
  Let $X =\VecteurLigneCoins{x_1}{x_N} \in L^N$.  It is clear that a
  linear combination with coefficients in $K$ is also a linear
  combination with coefficients in $L$, hence $w_1(x) \leq w_2(x)$.

  Let $ w_1 \defeq \rank_L \left( X^{\theta}
  \right)$. Noting the columns of $X^\theta$ 
\[
C_i={\VecteurLigneCoins{\Tordu{x_i}{0}}{\Tordu{x_i}{n-1}}}^T,
\]
suppose that the columns $C_{i_1},\ldots,C_{i_{w_1}}$ are
$L$-linearly independent. Then any $i$-th column can be written
$
C_i=\sum_{j=1}^{w_1}\lambda_jC_{i_j}
$, $\lambda_j\in  L$.
Applying $\theta^u$, we get
\[
\Tordu{C_i}{u}=\sum_{j=1}^{w_1}\Tordu{\lambda_j}{u} C_{i_j}^\theta, u=1,\ldots, m,
\]
which is the same as 
\[ C_i=\sum_{j=1}^{w_1}\Tordu{\lambda_j}{u} C_{i_j},  u=1,\ldots, m,
\]since $\Tordu{C_i}{u}$ is a cyclic shift of $C_i$. By summation, we get
\[
C_i=\sum_{j=1}^{w_1}\left(\sum_{u=0}^{m-1}\Tordu{\lambda_j}{u}\right)C_{i_j}.
\]
We have 
$$
 \left(\sum_{u=0}^{m-1}\Tordu{\lambda_j}{u}\right)^{\theta} = \sum_{u=1}^{m}\Tordu{\lambda_j}{u}.
$$
However $\theta$ has order $n$ which divides $m$. Therefore  $\Tordu{\lambda_j}{m} = \Tordu{\lambda_j}{0}$,
therefore $\sum_{u=0}^{m-1}\Tordu{\lambda_j}{u} \in K$ when $K=L^\theta$.
 This
implies that the columns $C_{i_1},\ldots,C_{i_{w_1}}$ $K$-generate the column space of
$X^\theta$: $w_2\leq w_1$.
\end{IEEEproof}
It is easy to see that the $w_i$'s provide distances defined by $
d_i(X,Y) \defeq w_i(X-Y)$.  In the following, we suppose
that we are in the case where all these metrics are equal, and the
induced distance is called rank metric. We use the notation $w(X)$,
without indices. This definition is a generalization of rank metric as
defined in Gabidulin~\cite{Gabidulin:1985}.

\begin{xpl}
Here is an example of a vector whose ranks are different on $K$ and on $L$.
Let us consider again the field extension
\[
K = \mathbb{Q} \hookrightarrow L=\mathbb{Q}[Y]/(Y^8+1).
\]
Let $\alpha$ be a root of $Y^8+1$, such that $(1,\ldots,\alpha^7)$ is a $K$-basis of $L$. Consider again the automorphism $\theta$ defined by $ \alpha \mapsto \alpha^3$. 
Let $x=(1,\alpha,\alpha^2,\alpha^4,\alpha^5,3\alpha^4+2)$. We have that $w_0(x)=w_1(x)=4 \leq w_2(x)=w_3(x)=5$
\end{xpl}

%%% Local Variables: 
%%% mode: latex
%%% TeX-master: "Soumission_ISIT"
%%% End: 

\section{Gabidulin codes in characteristic zero}

For simplicity, we suppose in this section that the  automorphism $\theta$ satisfies the following properties:
\begin{itemize}
\item $\theta$ generates the Galois group of $K \hookrightarrow L$,
  that is $\theta$ has order $m$;
\item The characteristic polynomial of $\theta$ is square-free;
\item $L^{\theta} = K$.
\end{itemize} 
The $K$-vector space $L^N$ is endowed with the rank metric defined in
the previous section.  In this metric space, a linear code is as usual
an $L$-vector space of length $N$, dimension $k$ and minimum rank
distance $d$.  It is denoted a $[N,k,d]_{(L,\theta)}$ code.

\subsection{Definition}

\begin{defi}
  Let $g=\VecteurLigneCoins{g_1}{g_N} \in L^N$, be $K$-linearly
  independent elements of $L$.  The generalized Gabidulin code, with
  dimension $k$ and length $N$, denoted $Gab_{\theta,k}(g)$, as a
  $L$-subspace of $L^N$, is $L$-generated by the matrix
$$G \defeq \MatriceCoins{\Tordu{g_1}{0}}{\Tordu{g_N}{0}}{\Tordu{g_1}{k-1}}{\Tordu{g_N}{k-1}},$$
\end{defi}
For $k \leq N$, the dimension of $Gab_{\theta,k}(g)$ is indeed $k$. We
can show that the parity-check matrix of $Gab_{\theta,k}(g)$ can be
given by
$$
H \defeq\MatriceCoins{\Tordu{h_1}{0}}{\Tordu{h_N}{0}}{\Tordu{h_1}{d-2}}{\Tordu{h_N}{d-2}},
$$
where $d = N-k+1$ for some $h_i \in L$ which are also $K$-linearly independent. 
%% Est ce que cela se montre avec la meme demonstration que Gabidulin ?
%%

\subsection{Maximum Rank Distance codes}

\begin{prop}
Let $\mathcal{C}$ be an $[N,k,d]_{(L,\theta)}$ code. We have $d\le N-k+1$.
\end{prop} 
\begin{IEEEproof}
Omitted due to lack of space.
\end{IEEEproof}

%% Pareil, est-ce que cela se montre comme dans le cas de Gabidulin ????
An optimal code satisfying the property that $d=N-k+1$ is called a 
Maximum Rank Distance (MRD) code.

\begin{thm}
The generalized Gabidulin $Gab_{\theta,k}(g)$ is an MRD code.
\end{thm}

\begin{IEEEproof}
Let $C = \VecteurLigneCoins{c_1}{c_N} \in Gab_{\theta,k}(g)$ be a non-zero codeword. By definition of generalized Gabidulin codes, 
there exists a $\theta$-polynomial $P(X)$ of $\theta$-degree $\le k-1$ such that 
\[
\forall i=1,\ldots,N,\quad c_i =  P(g_i).
\]
Now, $C$ has rank $d$ if and only if the $K$-vector space generated by its components has $K$-dimension $d$. Therefore, 
by Th. \ref{Thm:Lagrange}, there exists a $\theta$-polynomial of $\theta$-degree $d$ such that $P_C(c_i) = 0$ for all $i$.
Hence 
$$
\forall i = 1,\ldots, N, \quad P_C \times P (g_i) = 0.
$$ 
Since $<g_1,\ldots,g_N>$ has $K$-dimension $N$, since $P$ has degree at most $k$, and since we are in the case where the dimension of the 
root-space of a $\theta$-polynomial is at most its degree, we have   $d+k-1 \ge N$ therefore $d -1 = N-k$.  
\end{IEEEproof}

\subsection{Unique decoding}
Our version of the algorithm is inspired from Gemmel and Sudan's
presentation of the algorithm of
Welch-Berlekamp~\cite{Gemmell-Sudan:IPL1992}.  A more efficient
variant can be used using~\cite{Loidreau:2005}, but we prefer to
present here a more intuitive version.  Consider a vector
$Y=\VecteurLigneCoins{y}{y} \in L^N$ such that there exists
$E=\VecteurLigneCoins{e}{e}\in L^N$ such that
\begin{IEEEeqnarray}{rcl}\label{decodingsituation} Y = C + E,\\
C \in Gab_{\theta,k}(g),\\
\rank(E)\leq (N-k)/2.
\end{IEEEeqnarray}  
Write $t=\floor{(N-k)/2}$. We define the following series of problems
related to this situation.
\begin{defi}[Decoding]
Given $Y\in L^N$, find, if it exists, a pair $(f,E)$ such that
$y_i=f(g_i)+e_i$, $i=1,\ldots,N$ ; $w(E) \leq t$ ; $\deg_{\theta}(f) < k$.
%MODIF
%En fait, on ne s'interesse pas a retrouver C, ce qui nous interesse, ce sont uniquement les coefs de f.
%Non seulement C ne ne permet pas de retrouver l'information de départ (on a seulement oté l'erreur et il reste à faire une interpolation "classique"),
%mais en plus on utilise f pour le calculer.
%Given $Y\in L^N$, find, if it exists, a pair $(C,E)$ such that
% $C \in Gab_{\theta,k}(g)$, $w(E) \leq t$, $Y=C+E$.
\end{defi}
\begin{defi}[Nonlinear reconstruction]
  Given $Y\in L^N$, find, if it exists, a pair of $\theta$-polynomials
  $(V,f)$ such that $\deg_{\theta}(V) \leq t$ ; $V \neq 0$ ; 
  $\deg_{\theta}(f) < k$ ;  $ V(y_i) = V(f(g_i))$, $i=1,\ldots,N$.
\end{defi}
Note that this problem gives rise to quadratic equations, considering
as indeterminates the coefficients of the unknowns $(V,f)$ over the
basis $\mathcal{B}$. We thus consider a linear version of the system.
\begin{defi}[Linearized reconstruction]
  Given $Y\in L^N$, find, if it exists, a pair of $\theta$-polynomials
  $(W,N)$ such that
 $\deg_{\theta}(W) \leq t$ ; $W \neq 0$ ;
 $\deg_{\theta}(N) < k+t$ ;
 $W(y_i) = N(g_i)$, $i=1,\ldots,N$.
\end{defi}
Since we require the weight of the error to be less than or equal to
$t=(N-k)/2$, we have unicity of the solution for the three above
problems. Now the following propositions give relations between the
solutions of these problems.
\begin{prop}
  Any solution of \emph{Nonlinear reconstruction} give a solution of
  \emph{Decoding}.
\end{prop}
\begin{IEEEproof}
Let $(V,f)$ be a solution of \emph{Nonlinear reconstruction}.  We define
$e_i \defeq y_i-f(g_i)$. Then we have $y_i=f(g_i)+e_i$, $i=1,\ldots,N$ ; $\deg_{\theta}(f) < k$ ; $w(E) \leq t$.
Indeed, since the $e_i$'s are roots of a $\theta$-polynomial with degree at most $t$,we must have $\deg \min(I_E)\leq t$, thus, $w(E)\leq t$.
%MODIF Et on modifie cette preuve de la même façon.
%  Let $(V,f)$ be a solution of \emph{Nonlinear reconstruction}.  We
%  define $C=\VecteurLigneCoins{c_1}{c_N}$, with $c_i \stackrel{def}{=} f(g_i)$,
%  $E=\VecteurLigneCoins{e_1}{e_N}$, with $e_i \stackrel{def}{=} y_i-f(g_i)$, $i=1,\cdots,N$. Then
%  we have that $y_i=c_i+e_i$, $C \in Gab_{\theta,k}(g)$, and $w(e)
%  < t$. Indeed since the $e_i$'s are roots of a $\theta$-polynomial with
%  degree at most $t$, we must have $\deg \min(I_E)\leq t$, thus, $w(E)\leq t$.
\end{IEEEproof}
Under an existence condition, we have the following statement.
 \begin{prop}
  If $t\leq (N-k)/2$, and if there is a solution to \emph{Nonlinear reconstruction}, then 
  any solution of \emph{Linear reconstruction}
  gives a solution to \emph{Nonlinear reconstruction}.
\end{prop}
\begin{IEEEproof}
  Let $(V,f)$ be a non zero solution of \emph{Nonlinear
    reconstruction}, and let $(W,N)$ be a solution of \emph{Linearized
    reconstruction}. Letting $e_i \defeq y_i -f(g_i)$, $i=1,\ldots,N$, we
  have $V(e_i) = V(y_i-f(g_i)) = 0$. Thus $V\in I_E$, with $\deg V\leq
  t$, so $E=\VecteurLigneCoins{e_1}{e_N}$ has rank at most $t$.

  We also have $W(e_i)=W(y_i) - W(f(g_i))$ so $W(e_i) = N(g_i) -
  W(f(g_i))$. Since $W(e_i)$ has rank at most $t$, we can find $U$
  with degree at most $t$, such that $U(W(e_i)) = U(N(g_i) -
  W(f(g_i))) = 0$.

  Then $\left(U\times (N - W \times f)\right)(g_i) = 0$, $i=1,\ldots,N$.
  As $t \leq (N-k)/2$, degree computations show that $U \times (N - W
  \times f)$ is a $\theta$-polynomial with degree at most $N-1$. Since
  it is zero at $N$ $K$-linearly independent values, it must be the
  zero polynomial: $U\times (N - W \times f)=0$.  As there is no zero
  divisor in $L[X;\theta]$, we conclude that $N = W \times f$.  Then
  $(W,N) = (W,W \times f)$, and $(W,f)$ is a solution of \emph{Nonlinear
    reconstruction}.
\end{IEEEproof}
The above propositions imply that unique decoding is equivalent to 
solving \emph{Linearized reconstruction}.
 Now we give the explicit system of equations to be solved.
\begin{thm}
  Solving \emph{Linearized reconstruction} amounts to solving the
  following linear system of equations
  \[
S \cdot \begin{pmatrix} {N} \\ - {W} \end{pmatrix} = 0,
\]
where
\[ S \defeq\begin{pmatrix}
\Tordu{g_1}{0}  & \cdots & \Tordu{g_1}{k+t-1} & \Tordu{y_1}{0} & \cdots & \Tordu{y_1}{t} \\ 
\vdots          & \ddots & \vdots             & \vdots         & \ddots & \vdots         \\ 
\Tordu{g_N}{0}  & \cdots & \Tordu{g_N}{k+t-1} & \Tordu{y_N}{0} & \cdots & \Tordu{y_N}{t}
\end{pmatrix} \]
with unknowns
\begin{IEEEeqnarray*}{rll}
{N} &= \VecteurLigneCoins{n_0}{n_{k+t-1}}^T\in L^{k+t}\\
{W} &=  \VecteurLigneCoins{w_0}{w_{t}}^T\in L^{t+1}.
\end{IEEEeqnarray*}
\end{thm}
\vspace{1ex}
\begin{IEEEproof}
  Each row of the product corresponds to the evaluation of ${N}$ and
  ${W}$ in the $g_i$'s and in the $y_i$'s.
\end{IEEEproof}
\begin{rmq}
  The number of arithmetic operations used in this method is easily
  seen to be of $O(N^3)$, using for instance Gaussian elimination for
  solving the linear system. However, since the system is highly
  structured, a better algorithm exists~\cite{Loidreau:2005} whose
  complexity is $O(N^2)$.
\end{rmq}
\begin{rmq}
  Note that we only deal with the algebraic complexity, i.e.\ the
  number of elementary additions and multiplications in $L$. Since we
  may compute over infinite fields, this does not reflect the
  bit-complexity, which shall be studied in a longer version of the
  paper.
\end{rmq}
% \begin{prop}
%   In a context of decoding, the degrees of $W$ and $N$ are such that
%   the euclidean division give a polynomial $f$ with good degree. So it
%   is not necessary, in this context, to assume that $W$ has rank
%   exactly $t$.
% \end{prop}
% \vspace{1ex}
% \begin{IEEEproof}
% In a context of decoding, we know that exist $(e_i)$ and $(c_i)$ such that $y_i=c_i+e_i$, $c \in \mathcal{C}$ and 
% $rank(e) = \tilde{t} \leq t$.\\
% As $c \in \mathcal{C}$, we can write $c_i=f(g_i)$ where $f$ is a $\theta$-polynomial with degree at most $k-1$. We also have $V$ which cancel $(e_i)$. Choosing $V$ with minimal degree, it has rank $\tilde{t}$.\\
% We deduce a solution $(W,N) = (V,V \circ f)$ with respective degrees $\tilde{t}$ and $k-1+\tilde{t}$. So the quotient has degree $k-1$.\\
% We also deduce solutions with degrees from $(\tilde{t}, k-1+\tilde{t})$ to $(t,k-1+t)$ left-multiplying this solution by $\Tordu{X}{1}$.
% \end{IEEEproof}
% \vspace{1ex}
% \vspace{1ex}
% \begin{thm}[Decoding Algorithm]
% The decoding of generalized Gabidulin codes is as follows :
% \begin{itemize}
% \item We compute a solution $(W,N)$ of $Reconstruction2$
% \item We compute the division of $W$ by $N$ to get $f$ such that $N=W \circ f$
% \item So we have $y_i = f(g_i) + e_i$.
% \end{itemize}
% \end{thm}

%%% Local Variables: 
%%% mode: latex
%%% TeX-master: "Soumission_ISIT"
%%% End: 

\section{Examples}

We have previously seen the importance of the hypotheses about
$\theta$ and what happen when they are not satisfied.  Now, we will
see that Kummer extensions always provide automorphisms with the good
properties.

\begin{xpl}
Let us consider the Kummer extension
\[
K = \mathbb{Q}[X]/(X^4+1) \hookrightarrow L=K[Y]/(Y^8-3).
\]
Let $h$ be a root of $X^4+1$, such that $(1,h,h^2,h^3)$ is a $\mathbb{Q}$-basis of $K$, and let $\alpha$ be a root of $Y^8-3$, such that $(1,\ldots,\alpha^7)$ is a $K$-basis of $L$.
Consider this time the automorphism $\theta$ defined by $ \alpha \mapsto h \alpha$. Its characteristic polynomial is $Y^8-1$, which is square-free.
Thus, we can define generalized Gabidulin codes with symbols in $L$, of length $8$, and any dimension less than or equal to $8$. Besides being simply $\mathbb{Q}$-linear, these codes are also $K$-linear.
\end{xpl}
More generally, with Kummer extensions, we can design rank-metric
$[N,k,d]$ codes, accomplishing the MRD condition 
$N-k=d-1$. Below is also given a classical infinite family.

\begin{xpl}
Consider $p$ an odd prime number, and let $\zeta$ be a primitive
$p$-root of unity in $\C$. Then $\Q\hookrightarrow \Q[\zeta]$ is an
extension of degree $p-1$, and its Galois group is isomorphic to
$(\Z/pZ)^\star$, and is thus cyclic. We let $K=\Q$, and
$L=\Q[\zeta]$. For any $u$ with $\gcd(u,p-1)=1$, consider
$\theta:\zeta\mapsto\zeta^u$. Then $\theta$ has order $p-1$ and $\Q$
is the subfield stable under $\theta$. Then, for $k\leq p-1$, can build $\mathbb{Q}$-codes in
$L^{p-1}$, of dimension $k$ over $L$, such that the $K$-rank of any
codeword is at least $(p-1)-k+1=p-k$.
\end{xpl}

\section{Conclusion}
For a $\theta$-polynomial, we have seen the link between its degree
and the dimension of its kernel. Particularly, we gave sufficient
condition for the root-space dimension being at most the degree of a
$\theta$-polynomial, namely.

Then, we have seen four different ways to define notions related to
the rank-metric. This reduces to only two metrics, which are
furthermore the same in the case of $\theta$ having a square-free
characteristic polynomial. 

We have also given a generalized definition of Gabidulin codes, seen
that they are MRD codes, and can be easily decoded up to half the
minimum distance. Since computations are not carried over finite
fields, the bit complexity will be properly evaluated in the future.

Finally, properly applying this theory to space-time coding needs
further work.

% To finish, we have an example of extension and automorphism which have
% the good properties : if $K \hookrightarrow L$ is a Kummer extension
% with degree $d$, we have $\phi (d)$ valid automorphisms : any
% generator of the cyclic Galois group of the extension.

% Otherwise, we can reduce to this case defining $K_{int} := \{x \in L :
% \Tordu{x}{1}=x \}$ and considering the intermediate extension $K_{int}
% \hookrightarrow L$, whose Galois group is cyclic. Also, any of its
% generators has for eigenvalues all the roots of unity, so they have
% single multiplicity.

%%% Local Variables: 
%%% mode: latex
%%% TeX-master: "Soumission_ISIT"
%%% End: 

\bibliographystyle{plain}
\bibliography{isit}

\end{document}